\documentclass[a4paper,12pt]{iopart}

\usepackage{iopams}
\usepackage{graphicx}
\usepackage{booktabs}
\usepackage{amsmath,amssymb,amsfonts}
\usepackage{xcolor}
\usepackage{hyperref}
\IfFileExists{orcidlink.sty}{\usepackage{orcidlink}}{\providecommand{\orcidlink}[1]{}}
\IfFileExists{placeins.sty}{\usepackage[section]{placeins}}{}

\hypersetup{colorlinks=true, linkcolor=blue, citecolor=blue, urlcolor=blue,
            breaklinks=true}

\providecommand{\MLST}{Mach.\ Learn.: Sci.\ Technol.}

\providecommand{\norm}[1]{\left\lVert#1\right\rVert}
\newcommand{\Var}{\mathrm{Var}}
\newcommand{\Zexp}[1]{\langle Z_{#1}\rangle}

\begin{document}

\title[Capacity ceiling, not barren plateau]
      {An architectural capacity ceiling, not a barren plateau:
       why a fixed-encoding variational quantum circuit cannot fit the
       Lorenz-63 attractor}

\author{Tushar Pandey\,\orcidlink{0000-0001-7448-5723}}

\address{Texas A\&M University, College Station, TX 77843, USA}

\ead{tusharp@tamu.edu}

\submitto{\MLST}

\begin{abstract}
Variational quantum circuits train poorly on chaotic forecasting, a failure
usually blamed on barren plateaus (exponentially vanishing gradients). Using an
exactly simulable four-qubit variational quantum physics-informed circuit fit to
the Lorenz-63 system, we show the barren-plateau explanation fails and identify
the failure as an architectural capacity ceiling fixed by the circuit's
time-encoding, not its trainable depth. Four measurements support this. (i)~A
McClean-comparable gradient-variance estimator sits at the local-cost
Haar/2-design scale $2^{-2n}=3.9\times10^{-3}$ at $n=4$; on its structurally live
parameters it decays about ninefold with depth, then saturates there, large
enough to train, not an exponential collapse to zero. (ii)~At a common nominal
budget of $200$ optimiser iterations ($600$, in three stages, for layer-wise),
three optimiser families (gradient descent, layer-wise, SPSA) reach the same
order of magnitude of loss, with finite-difference and exact parameter-shift
gradients agreeing to about $2\times10^{-5}$ relative as a consistency check, so
no optimiser unlocks a better basin. (iii)~The output-Jacobian rank saturates at
$33$ from five layers on, so depth buys no new output directions. (iv)~A Fourier
analysis explains why: the qubit-1 phase encoding acts on the initial
$|0\rangle$ and is inert, so the maximum accessible frequency is
$2.5/t_{\max}\approx0.83$~Hz, identical at every depth and a factor of about
$4.4$ below the narrowest Lorenz component's bandwidth. The corrected band has
dimension $1+2\times5=11$ per observable, and $3\times11=33$ equals the measured
rank ceiling exactly, unifying the two diagnostics. A trained depth sweep
agrees: mean loss improves with depth, then flattens once the rank saturates. We
correct our earlier preprint diagnosis, which compared unnormalised gradient
norms to the McClean threshold, and place the advantage of fixed reservoirs and
classical echo-state networks in architecture, not quantum mechanics.
\end{abstract}

\vspace{2pc}


\section{Introduction}

A recurring empirical observation in quantum machine learning is that
variational quantum circuits, trained end-to-end by gradient descent, struggle
to fit hard temporal targets such as chaotic
trajectories~\cite{mcclean2018,cerezo2021,pandey2604}. When such a model fails
to reach a useful loss, the default explanation in the literature is the barren
plateau: for sufficiently expressive random circuits, the variance of the loss
gradient with respect to any single parameter vanishes exponentially in the
qubit count or the circuit depth, so that the loss landscape becomes an
exponentially flat plateau on which gradient descent stalls~\cite{mcclean2018,
cerezo2021}.

Testing that diagnosis requires care about scale. At fixed qubit count an
absolute-threshold statement is not available: the McClean bound is a scaling
statement in the qubit number $n$, and a single small $n$ cannot be used to
assert that a variance has or has not fallen below an $n$-asymptotic threshold.
What a single $n$ can show is whether the gradient variance decays with circuit
size and where it sits relative to the Haar/2-design saturation scale for the
same ansatz and cost. That comparison must be made with a normalised,
dimensionless quantity: an absolute gradient magnitude carries the arbitrary
scale of the loss function and cannot be compared to a $2^{-n}$ threshold at
all. Neglecting this normalisation is precisely the error we correct here
relative to our own earlier preprint (Section~\ref{ssec:walkback}).

This paper studies a deliberately small and exactly simulable case: a four-qubit
variational quantum physics-informed neural network (QPINN) asked to fit the
Lorenz-63 attractor~\cite{lorenz1963,raissi2019,qcpinn2025,leung2022qpinn}.
Small scale is deliberate. At four qubits every diagnostic we need can be
computed exactly, with exact parameter-shift gradients and full statevector
Jacobians, so no statistical or sampling artefact can be mistaken for a barren
plateau. We make no claim about near-term hardware relevance; the contribution
is a mechanistic diagnosis, not a device demonstration.

\paragraph{Contribution.}
We show that the QPINN failure on Lorenz-63 is not a barren plateau but an
architectural capacity ceiling set by the circuit's fixed time-encoding. The
evidence is four measurements (Section~\ref{sec:results}):
\begin{enumerate}
  \item a McClean-comparable gradient-variance estimator that sits at the
        local-cost Haar/2-design floor and, on its live parameters, decays
        toward that floor with depth rather than collapsing to zero
        (Section~\ref{ssec:variance});
  \item optimiser-independence of the final loss at matched iteration budget
        (Section~\ref{ssec:optimiser});
  \item saturation of the output-Jacobian effective dimension with depth
        (Section~\ref{ssec:jacobian});
  \item a depth-independent Fourier-expressivity ceiling below the target
        bandwidth (Section~\ref{ssec:fourier}), whose band dimension equals
        the measured rank ceiling exactly (Section~\ref{ssec:closure}),
\end{enumerate}
consistent with a trained depth sweep whose mean loss is flat once the Jacobian
rank has saturated (Section~\ref{ssec:capcurve}). We then correct our earlier
gradient-norm diagnosis (Section~\ref{ssec:walkback}) and argue that the
observed advantage of a fixed reservoir, and of a classical echo-state network,
over the variational circuit is architectural, not quantum
(Section~\ref{ssec:architectural}).

\section{Related work}

\paragraph{Barren plateaus and trainability.}
McClean et al.~\cite{mcclean2018} established that random parameterised quantum
circuits exhibit exponentially concentrating expectation values and hence
exponentially small gradient variance, and Cerezo et al.~\cite{cerezo2021}
surveyed the resulting trainability landscape and its distinct causes
(expressibility-induced plateaus, cost-function locality, entanglement-induced
plateaus, and noise-induced plateaus). Cost-function locality matters here: for
a local cost the Haar/2-design variance floor is $\Theta(2^{-2n})$ rather than
$\Theta(2^{-n})$~\cite{cerezo2021costlocality}, and our diagnostic cost is
local. Beyond the original global-cost analysis, the modern literature
distinguishes several trainability phenomena, including local barren plateaus,
layer-wise trainability, and observable-dependent
concentration~\cite{cerezo2021costlocality,uvarov2021locality,ortiz2021entanglement}.
Our live/dead parameter split (Section~\ref{ssec:methods-variance}) is exactly
an observable-locality analysis of this kind: the local cost reads only
$\Zexp{0}$, so at shallow depth many parameters sit outside the backward light
cone of $Z_0$ and have identically zero gradient, and we report the live
variance separately rather than diluting it with these structural zeros. On the
capacity side, expressivity- and approximation-theoretic tools have been used to
enlarge the reachable function class, for example tensor-network-guided
hypernetworks for variational circuits~\cite{qi2026tensorhyper}; our diagnosis
is complementary, showing that for a fixed encoding the reachable class is
band-limited independently of the trainable block. A central point of the
present paper is that these trainability pathologies are diagnostically distinct
from a capacity shortfall, and that the two must be separated with a properly
normalised measurement.

\paragraph{Expressivity of encodings.}
Schuld, Sweke, and Meyer~\cite{schuld2021effect} showed that a data-encoding
quantum circuit realises a truncated Fourier series in the input, whose
accessible frequency spectrum is fixed by the data-encoding gates, while the
trainable blocks only set the Fourier coefficients within that spectrum. Our
Fourier diagnostic (Section~\ref{ssec:fourier}) is a direct application of this
result: the maximum accessible frequency is a property of the fixed encoding
and is therefore independent of variational depth.

\paragraph{Reservoir computing, classical and quantum.}
Reservoir computing~\cite{jaeger2004harnessing,pathak2018model} fixes a
high-dimensional nonlinear dynamical system and trains only a linear readout,
and quantum reservoir
computing~\cite{fujii2017,mujal2021,nakajima2019boosting,ahmed2025robust}
transplants this to a fixed quantum map. Our companion fair-baseline
benchmarking study~\cite{pandey2607b} shows that on chaotic
forecasting a tuned classical echo-state network matches or beats a quantum
reservoir at comparable feature scale, and a companion conditioning
analysis~\cite{pandey2607} traces the reservoir advantage to the
bounded condition number of a fixed random feature map relative to an
end-to-end-trained one. The present paper is the trainability-side complement:
it explains why the variational alternative, specifically, is capacity-limited
on this task.

\paragraph{Relation to our prior work.}
The base QPINN/QRC architecture and the Lorenz forecasting setup were introduced
in~\cite{pandey2604}, which also reports the QPINN-versus-reservoir performance
gap we build on here; we reuse that setup and do not re-derive the circuits
beyond what is needed for self-containment. The diagnosis reported here
supersedes the trainability claim of an earlier, unpublished version of this
manuscript, whose gradient-norm comparison was not normalised to the McClean
threshold (Section~\ref{ssec:walkback}).

\section{Methods}
\label{sec:methods}

\subsection{Circuit and task}
\label{ssec:methods-circuit}

We reuse the four-qubit QPINN and the Lorenz-63 forecasting task
of~\cite{pandey2604}, summarised here for self-containment. Time is normalised
to $[0,2\pi]$ and encoded by single-qubit rotations,
\begin{equation}
  \theta_t = \frac{2\pi t}{t_{\max}}, \qquad
  U_{\mathrm{enc}}(t) = R_Y^{(0)}(\theta_t)\,R_Z^{(1)}(\theta_t)\,
                        R_X^{(2)}(\theta_t)\,R_Y^{(3)}(\theta_t/2),
\end{equation}
where the fourth qubit is driven at half the base frequency. Each of $L$
variational layers applies $R_X R_Y R_Z$ rotations on every qubit followed by a
CNOT entangling chain with $R_Z$ phases on the targets, giving $15L$ trainable
parameters. The physical state estimate is a linear map of the Pauli-$Z$
expectations $\Zexp{0},\Zexp{1},\Zexp{2}$. The physics-informed loss combines
the Lorenz ODE residual at $N_c=50$ collocation times with an initial-condition
term,
\begin{equation}
  \mathcal{L}_\theta
    = \frac{1}{N_c}\sum_{i=1}^{N_c}
        \norm{\partial_t f_\theta(t_i) - F(f_\theta(t_i))}^2
    + \lambda\,\norm{f_\theta(0) - x_0}^2 ,
\end{equation}
with $\lambda=10$ and $F$ the Lorenz vector field
($\sigma=10,\rho=28,\beta=8/3$). All quantities below are computed by exact
statevector simulation on $n=4$ qubits.

One structural feature of this encoding is load-bearing for the Fourier
diagnostic and is worth stating up front. The qubit-1 encoding gate
$R_Z^{(1)}(\theta_t)$ is the first gate acting on that qubit, which starts in
$|0\rangle$; an $R_Z$ rotation on a computational-basis state contributes only a
global phase, so the qubit-1 encoding frequency is inert and never enters any
observable. We confirm this numerically in Section~\ref{ssec:fourier}: deleting
the gate changes $\Zexp{0},\Zexp{1},\Zexp{2}$ by at most $1.6\times10^{-15}$ at
every depth. The consequence is that the accessible spectrum is set by three
active encoding frequencies, not four (Section~\ref{ssec:methods-fourier}).

\subsection{McClean-comparable gradient variance}
\label{ssec:methods-variance}

The barren-plateau prediction is a statement about the variance, across random
initialisations, of the gradient of a fixed circuit (a fixed input) with respect
to a single parameter. It is essential that no averaging over the input (here,
over time) is performed before differentiating, because time-averaging the cost
annihilates all non-DC Fourier modes of the encoding and destroys exactly the
quantity the McClean bound concerns.

We define the McClean-comparable estimator on the bounded, local cost
$C_t = \tfrac{1}{2}\bigl(1 - \Zexp{0}(t)\bigr) \in [0,1]$ at each fixed
collocation time $t$. For each of $100$ random initialisations we compute the
exact gradient $\partial C_t/\partial\theta_k$ by the two-evaluation
$\pm\pi/2$ parameter-shift rule (no finite differencing). Because each $t$ is a
fixed circuit and the cost is bounded in $[0,1]$, the across-initialisation
variance $\Var_{\text{init}}[\partial_{\theta_k}C_t]$ is dimensionless and
directly comparable to a $2^{-n}$-type scale. We sweep $L\in\{1,3,5\}$ (that is,
$15,45,75$ parameters) on $24$ fixed collocation times.

The cost $C_t$ reads only $\Zexp{0}$, and the CNOT entangler flows one-way along
$0\!\to\!1\!\to\!2\!\to\!3$. In the Heisenberg picture $Z_0$ therefore commutes
with the entire entangler and with every last-layer gate except $R_X^{(0)},
R_Y^{(0)}$, so at low depth a large fraction of parameters have identically zero
gradient: they lie outside the backward light cone of $Z_0$. We report the
pooled mean variance $\Var_{\text{est}}=\langle
\Var_{\text{init}}[\partial_{\theta_k}C_t]\rangle_{k,t}$, but we also report it
separately for the structurally live and dead parameters, because pooling over a
dead-parameter fraction that shrinks with depth confounds light-cone geometry
with plateau physics. A parameter-time cell is counted dead when its
across-initialisation variance is below $10^{-28}$; the maximum absolute
gradient among dead cells is $3.5\times10^{-16}$, confirming these are exact
structural zeros rather than small finite values.

\subsection{Matched-budget optimiser comparison}
\label{ssec:methods-opt}

To test whether the failure is an artefact of a single optimiser, we train the
full physics loss with four training configurations at a common nominal budget of $200$
optimiser iterations (with the layer-wise and SPSA exceptions detailed in the
protocol caveats below): (a) forward finite-difference gradients ($\epsilon=10^{-4}$,
$10$ seeds); (b) exact parameter-shift gradients propagated analytically through
the quadratic loss ($5$ seeds); (c) a layer-wise schedule that optimises one
variational layer at a time ($5$ seeds); and (d) SPSA ($5$ seeds). Sources
(a)--(c) share the Adam meta-optimiser settings, learning rate, decay, and
gradient-clipping percentile of~\cite{pandey2604}; SPSA instead uses standard
Spall gain schedules ($\alpha=0.602$, $\gamma=0.101$, median-calibrated step
size) with no Adam and no clipping.

Three protocol details are needed to read Table~\ref{tab:optimiser} honestly and
are stated once here. First, the layer-wise schedule runs $200$ iterations per
stage for three stages, i.e.\ $600$ Adam updates per seed, three times the
per-optimiser iteration count of the others; its being the worst performer with
the most updates strengthens, not weakens, the optimiser-independence
conclusion. Second, SPSA was run with a gain schedule calibrated for a
$2000$-iteration horizon, and we read its exact loss after $200$ updates from
the logged \texttt{loss\_exact} field (the record at iteration $201$, since
\texttt{loss\_exact} is evaluated before each update and logged every ten
iterations); the same run reaches $1.27\times10^{3}$ by iteration $2000$, which
we report in the text so the reader can see the $200$-iteration runs are not
converged. Third, iteration-matching is not compute-matching: at $L=3$ the
per-iteration loss-evaluation counts differ, giving totals of about $9{,}200$
(finite-difference), $9{,}600$ (layer-wise), $36{,}200$ (parameter-shift, which
evaluates a three-point time-stencil Jacobian each iteration), and $\sim\!425$
(SPSA at iteration $200$). Only finite-difference and layer-wise are matched in
evaluations (to $1.04\times$); the comparison is on a common per-iteration
budget (three stages of it for layer-wise), and we make no compute-matched
claim.

Optimiser (b) uses the parameter-shift Jacobian of the observables composed with
the analytic Jacobian of the loss, and was verified against central finite
differences. In exact simulation forward finite differencing ($\epsilon=10^{-4}$)
and exact parameter-shift feed the identical Adam loop and reproduce the same
gradient, so their per-seed final losses agree to about $2\times10^{-5}$
relative on all shared seeds; rows (a) and (b) are therefore a single gradient
source measured twice, a consistency check that no finite-difference artefact is
present, not two independent optimisers. The honest count is three optimiser
families: gradient descent (a/b), block-coordinate layer-wise (c), and SPSA (d).

\subsection{Jacobian effective dimension}
\label{ssec:methods-jac}

For a fixed initialisation we form the exact output Jacobian
$J = \partial(\Zexp{0},\Zexp{1},\Zexp{2})(t)/\partial\theta$ at each of $50$
collocation times by parameter-shift, stack the per-time Jacobians, and take the
singular-value spectrum $\{s_i\}$ of the stacked matrix. We report two
depth-resolved effective-dimension measures, each averaged over $6$ random
initialisations: the numerical rank at relative threshold $10^{-6}$, and the
participation ratio $\mathrm{PR} = (\sum_i s_i^2)^2 / \sum_i s_i^4$, which counts
the number of singular directions carrying appreciable weight. We sweep
$L\in\{1,\dots,8\}$ (that is, $15$ to $120$ parameters).

\subsection{Fourier-expressivity ceiling}
\label{ssec:methods-fourier}

Following Schuld et al.~\cite{schuld2021effect}, a time-encoding gate
$R(\phi(t))$ with $\phi=\omega t$ and generator eigenvalues $\pm\tfrac12$
contributes frequencies $\{-f,0,+f\}$ with $f=\omega/2\pi$. The model's
accessible spectrum is spanned by integer sums and differences of the active
per-gate frequencies, so its maximum accessible frequency is their sum. Of the
four encoding gates, the qubit-1 $R_Z$ is inert (Section~\ref{ssec:methods-circuit};
verified numerically in Section~\ref{ssec:fourier}), leaving three active
frequencies: two qubits at $f=1/t_{\max}$ (qubits $0$ and $2$) and one at
$f=1/2t_{\max}$ (qubit $3$), with $t_{\max}=3$~s. The accessible harmonics are
therefore multiples of $1/2t_{\max}$ up to
\begin{equation}
  f_{\max} = \frac{1+1+\tfrac12}{t_{\max}} = \frac{2.5}{t_{\max}}
           \approx 0.83~\text{Hz},
  \label{eq:fmax}
\end{equation}
that is, the set $\{0,\tfrac12,1,\tfrac32,2,\tfrac52\}/t_{\max}$ of $1$ DC plus
$5$ non-zero harmonics. This is an analytic property of the fixed encoding and
is independent of the trainable ansatz. To verify it numerically rather than
recite an analytic constant, we sample $\Zexp{j}(t)$ over a full period
$[0,2t_{\max})$ (so every accessible half-integer harmonic lands on an exact FFT
bin, avoiding the spectral leakage that contaminates a half-period window) and
measure the relative spectral power above $f_{\max}$ at each depth. We compare
$f_{\max}$ to the $99$th-percentile spectral bandwidth of the reference Lorenz
trajectory, computed by FFT of each component of an RK4 solution at $\Delta t=0.01$.

\subsection{Trained depth sweep}
\label{ssec:methods-capcurve}

Finally, we train the full physics loss at a matched budget ($800$ SPSA
iterations, $3$ seeds each) across $L\in\{1,2,3,5,8\}$ and report the mean and
standard error of the final loss, to test whether the empirically achievable
loss tracks the structural capacity ceiling identified by the Jacobian and
Fourier diagnostics.

\subsection{Reproducibility}

All diagnostics are produced by a single checkpointed harness
(\texttt{scripts/r1\_experiments.py}) with recorded git revision, source-file
SHA-256, and per-unit checkpoints; every result file carries its provenance
block. The corrected live/dead variance split, capacity-curve dispersion, and
optimiser evaluation counts are computed post hoc from those files by
\texttt{scripts/analyze\_r1\_corrections.py}, and the corrected Fourier ceiling
is verified by an independent statevector simulation in
\texttt{scripts/verify\_fourier\_ceiling.py}; both are deterministic and carry
their own SHA-256. Result files are named in the captions below.

\section{Results}
\label{sec:results}

\subsection{The gradient variance sits at the Haar floor, not on a plateau}
\label{ssec:variance}

\begin{figure}[htbp]
  \centering
  \includegraphics[width=0.9\linewidth]{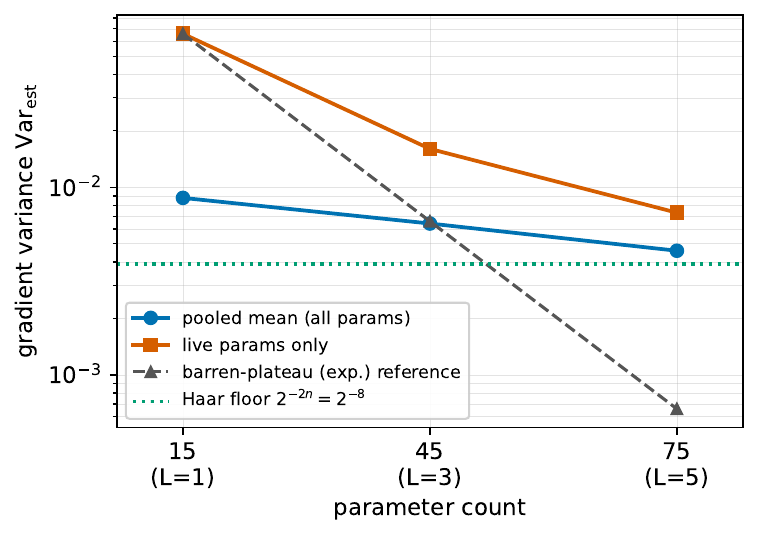}
  \caption{McClean-comparable gradient variance versus parameter count (log
    scale). The pooled mean (blue circles) is nearly flat, but this is a pooling
    artefact; the live-parameter mean (orange squares) decays about ninefold
    toward the local-cost Haar/2-design floor $2^{-2n}=3.9\times10^{-3}$ (green
    dotted line) and saturates there. A barren plateau would instead fall
    exponentially (grey dashed reference).
    Source: \texttt{variance\_b1\_e1\_L\{1,3,5\}.json}.}
  \label{fig:variance}
\end{figure}

Table~\ref{tab:variance} reports the McClean-comparable estimator of
Section~\ref{ssec:methods-variance}, split into pooled, live, and dead
parameters. Three facts follow. First, the pooled mean is essentially flat as
the parameter count grows fivefold ($8.8\times10^{-3}$ at $L=1$ to
$4.6\times10^{-3}$ at $L=5$, a factor of $1.9$); taken alone this looks like the
absence of any depth trend. Second, that flatness is a pooling artefact: the
dead-parameter fraction falls from $86.7\%$ ($L=1$) to $60.0\%$ ($L=3$) to
$37.3\%$ ($L=5$) as the backward light cone of $Z_0$ widens with depth, and the
pooled mean is exactly the product of the live-only mean and the live fraction.
Restricted to the live parameters, the variance decays about ninefold
($6.6\times10^{-2}\to7.3\times10^{-3}$, a factor of $9.0$). Third, and
decisively, that decay is toward a floor, not toward zero: for this local cost
the Haar/2-design saturation scale is $2^{-2n}=2^{-8}=3.9\times10^{-3}$ at
$n=4$~\cite{cerezo2021costlocality}, and the live variance falls from $16.9$ to
$1.9$ times this floor over $L=1$ to $L=5$. This is the canonical
decay-then-saturate profile of a fixed-$n$ circuit converging to its $2$-design
floor, and the floor itself is large enough to train. The failure is therefore
not a vanished gradient. Figure~\ref{fig:variance} plots the pooled and live
variances against the Haar floor and against the exponential decay a barren
plateau would predict.

We emphasise what a single $n$ can and cannot show. It cannot certify or refute
the $n$-asymptotic McClean threshold, which is inaccessible at $n=4$. It can show
that the gradient variance sits at the $2$-design floor for this ansatz and cost,
and that training reduced the loss tenfold as depth grew (Section~\ref{ssec:capcurve}),
so gradient magnitude was demonstrably not the binding constraint. The
defensible statement is that gradients at $n=4$ are at the Haar scale and are
large enough to train, not that they fail to decay with depth.

\begin{table}[htbp]
  \centering
  \caption{McClean-comparable gradient variance
    (Section~\ref{ssec:methods-variance}): the variance across $100$ random
    initialisations of the exact fixed-circuit, fixed-$t$ parameter-shift
    gradient of the bounded local cost, on $24$ collocation times. Pooling over
    a dead-parameter fraction that shrinks with depth makes the pooled mean look
    flat; the live-only mean decays about ninefold, from $16.9$ to $1.9$ times
    the local-cost Haar floor $2^{-2n}=3.9\times10^{-3}$. \textbf{Source:}
    \texttt{results/r1/variance\_b1\_e1\_L\{1,3,5\}.json} and
    \texttt{results/r1/r1\_corrections\_analysis.json} (\texttt{e1\_live\_dead}).}
  \label{tab:variance}
  \begin{tabular}{lcccc}
    \toprule
    $L$ & Dead frac. & Pooled mean & Live mean & Live / floor \\
    \midrule
    $1$ & $86.7\%$ & $8.8\times10^{-3}$ & $6.6\times10^{-2}$ & $16.9$ \\
    $3$ & $60.0\%$ & $6.4\times10^{-3}$ & $1.6\times10^{-2}$ & $4.1$ \\
    $5$ & $37.3\%$ & $4.6\times10^{-3}$ & $7.3\times10^{-3}$ & $1.9$ \\
    \bottomrule
  \end{tabular}
\end{table}

\subsection{The failure is optimiser-independent}
\label{ssec:optimiser}

If the poor loss were an optimiser pathology, a plateau that a better gradient
estimator could escape, then different optimisers should reach qualitatively
different basins. They do not. Table~\ref{tab:optimiser} shows that at a common
nominal budget of $200$ optimiser iterations ($600$, in three stages, for
layer-wise), gradient descent, layer-wise, and SPSA optimisation all
terminate within the same order of magnitude of final loss
($\sim\!10^{3}$--$10^{4}$). Finite-difference and exact parameter-shift agree to
about $2\times10^{-5}$ relative per seed (their row difference is entirely the
extra five finite-difference seeds), so they corroborate that no finite-difference
artefact is present rather than adding a fourth independent optimiser. The
layer-wise schedule, which runs three times as many Adam updates as the others
($600$ against $200$), is the worst performer, which sharpens the conclusion. No
optimiser unlocks a qualitatively better solution, consistent with a capacity
ceiling and inconsistent with a gradient-estimation problem.

The one caveat is convergence: SPSA continued to $2000$ iterations in the same
run reaches $1.27\times10^{3}$, so the $200$-iteration comparison establishes
that the optimisers agree at a common budget, not that any of them has reached a
basin floor. The qualitative conclusion (no optimiser unlocks a different order
of magnitude) is unaffected, and the Jacobian and Fourier diagnostics below give
the depth-resolved capacity limit directly.

\begin{table}[htbp]
  \centering
  \caption{Mean final loss at a common nominal budget of $200$ optimiser
    iterations on the $L=3$
    ($45$-parameter) QPINN. All land within one order of magnitude. Layer-wise
    runs $200$ iterations per stage for three stages ($600$ Adam updates);
    finite-difference and parameter-shift agree to about $2\times10^{-5}$
    relative and
    are one gradient source measured twice; SPSA is read after $200$ updates of
    a $2000$-iteration-scheduled run (record iteration $201$). Eval.\ counts are
    total loss evaluations at $L=3$ (loss-evaluation equivalents for
    parameter-shift, per the harness cost model) and show the comparison is
    iteration-matched, not compute-matched. \textbf{Sources:}
    \texttt{train\_fd\_b1\_e4\_fd\_10seed.json},
    \texttt{train\_ps\_b1\_e2b\_ps.json},
    \texttt{train\_layerwise\_b1\_e2d\_layerwise.json} (\texttt{final\_loss}),
    \texttt{train\_spsa\_b1\_e2a\_spsa.json}
    (\texttt{loss\_exact} at iteration $201$).}
  \label{tab:optimiser}
  \begin{tabular}{lccc}
    \toprule
    Optimiser & Seeds & Mean final loss & Evals \\
    \midrule
    Finite difference        & $10$ & $1.02\times10^{4}$ & $9{,}200$ \\
    Parameter-shift (exact)  & $5$  & $9.51\times10^{3}$ & $36{,}200$ \\
    Layer-wise ($3$ stages)  & $5$  & $1.10\times10^{4}$ & $9{,}600$ \\
    SPSA (iter.\ $200$)      & $5$  & $3.1\times10^{3}$  & $425$ \\
    \bottomrule
  \end{tabular}
\end{table}

\subsection{The output-Jacobian effective dimension saturates with depth}
\label{ssec:jacobian}

Table~\ref{tab:jacobian} reports the effective dimension of the output Jacobian
as a function of depth. The numerical rank rises through $5,15,25$ at
$L=1,2,3$, reaches $32.7$ at $L=4$ (four of six initialisations give $33$, two
give $32$), and is then flat at exactly $33$ across all initialisations from
$L=5$ through $L=8$: adding the $45$ parameters from $L=5$ to $L=8$ produces zero
new independent output directions. The participation ratio continues to rise
($6.8\to9.2$ over $L=4$ to $L=8$), so the added parameters redistribute weight
among directions that already exist without creating new ones. This is a
directly measured capacity ceiling in the space of functions the circuit's
output can represent. Section~\ref{ssec:closure} shows that the ceiling value
$33$ is exactly the dimension of the corrected band-limited function space.

\begin{table}[htbp]
  \centering
  \caption{Effective dimension of the output Jacobian
    $\partial(\Zexp{0},\Zexp{1},\Zexp{2})/\partial\theta$ versus depth, averaged
    over $6$ initialisations. The numerical rank saturates at $33$ from $L=5$
    onward (mean $32.7$ at $L=4$) while the parameter count keeps growing.
    \textbf{Source:} \texttt{results/r1/jacobian\_b1\_e3\_jac\_L\{1..8\}.json},
    per-entry \texttt{rank\_1e-6} and \texttt{participation\_ratio}.}
  \label{tab:jacobian}
  \begin{tabular}{lccc}
    \toprule
    $L$ & Params & Rank ($10^{-6}$) & Participation ratio \\
    \midrule
    $1$ & $15$  & $5$    & $2.0$ \\
    $2$ & $30$  & $15$   & $3.5$ \\
    $3$ & $45$  & $25$   & $6.2$ \\
    $4$ & $60$  & $32.7$ & $6.8$ \\
    $5$ & $75$  & $33$   & $7.7$ \\
    $6$ & $90$  & $33$   & $8.2$ \\
    $7$ & $105$ & $33$   & $8.8$ \\
    $8$ & $120$ & $33$   & $9.2$ \\
    \bottomrule
  \end{tabular}
\end{table}

\subsection{A depth-independent Fourier-expressivity ceiling}
\label{ssec:fourier}

Why does the Jacobian rank saturate? Because the circuit is structurally
band-limited by its encoding. The accessible maximum frequency of
Equation~\eqref{eq:fmax} is $f_{\max}=2.5/t_{\max}\approx0.83$~Hz, set by the
three active encoding gates and independent of the trainable ansatz. We verify
this numerically rather than reciting an analytic constant. First, deleting the
inert qubit-1 $R_Z$ encoding changes $\Zexp{0},\Zexp{1},\Zexp{2}$ by at most
$1.6\times10^{-15}$ at every depth $L\in\{1,3,5,8\}$, confirming it contributes
no frequency. Second, sampling the observables over the full period
$[0,2t_{\max})$ and taking the FFT on the commensurate grid, the relative
spectral power above $f_{\max}$ is at machine precision
($\sim\!10^{-30}$) for every observable and every depth; nothing lies above the
ceiling. (An earlier random-draw spectrum sampled on a half-period window showed
apparent power above the ceiling; that was spectral leakage from the
half-integer harmonics falling between FFT bins, and it vanishes on the
full-period grid.) The band edge $f_{\max}$ itself becomes populated once depth
is sufficient to route the qubit-3 half-frequency through the entangler onto the
$Z_0$ cost observable, which is exactly the depth dependence the rank column
records.

Table~\ref{tab:fourier} compares the ceiling to the Lorenz target bandwidths.
Even the narrowest component (the $x$-component, $99$th-percentile bandwidth
$3.65$~Hz) exceeds the accessible ceiling by a factor of $4.4$; the $y$ and $z$
components ($4.32$ and $15.61$~Hz) exceed it by factors of $5.2$ and $18.7$.
Expressed as target spectral energy, $36\%$ of the $x$-component's energy,
$54\%$ of the $y$-component's, and $85\%$ of the $z$-component's lies above the
ceiling and is therefore unreachable by the circuit at any depth. Adding
variational layers reparametrises energy within the accessible band without
widening it, which is the mechanistic cause of the Jacobian-rank saturation in
Section~\ref{ssec:jacobian}: more depth cannot create output directions at
frequencies the encoding cannot reach.

\begin{table}[htbp]
  \centering
  \caption{Maximum accessible output frequency and the Lorenz target bandwidths.
    The accessible frequency is set by the fixed encoding, is independent of
    depth, and lies a factor of $4.4$--$18.7$ below the per-component target
    bandwidths; the final column is the fraction of that component's spectral
    energy above the ceiling. \textbf{Source:}
    \texttt{results/r1/fourier\_ceiling\_verification.json} (ceiling, depth
    independence) and
    \texttt{results/r1/fourier\_b1\_e3\_fourier\_L1.json}
    (\texttt{target\_bandwidth\_99pct\_hz}, \texttt{target\_spectrum}).}
  \label{tab:fourier}
  \begin{tabular}{lccc}
    \toprule
    Component & Target bw.\ (Hz) & Gap to $f_{\max}$ & Energy above \\
    \midrule
    $x$ & $3.65$  & $4.4\times$  & $36\%$ \\
    $y$ & $4.32$  & $5.2\times$  & $54\%$ \\
    $z$ & $15.61$ & $18.7\times$ & $85\%$ \\
    \midrule
    \multicolumn{4}{l}{$f_{\max} = 2.5/t_{\max} \approx 0.83$~Hz, identical at
      $L\in\{1,3,5,8\}$.} \\
    \bottomrule
  \end{tabular}
\end{table}

\subsection{The band dimension equals the measured rank ceiling}
\label{ssec:closure}

The Fourier ceiling and the Jacobian rank are the same result seen two ways. The
accessible band is $\{0,\tfrac12,1,\tfrac32,2,\tfrac52\}/t_{\max}$: one DC
component plus five non-zero harmonics, each contributing a cosine and a sine, so
$1+2\times5=11$ real basis functions per observable. With three observables
$\Zexp{0},\Zexp{1},\Zexp{2}$ the reachable output functions span at most
\begin{equation}
  3\times(1 + 2\times5) = 3\times 11 = 33
\end{equation}
dimensions, which is exactly the measured rank ceiling of
Table~\ref{tab:jacobian}. Two independent diagnostics, an analytic
encoding-frequency count and a numerical singular-value rank, therefore agree on
the single number $33$. This closure is the strongest single result of the
paper: it identifies the capacity ceiling with a countable set of accessible
Fourier modes and confirms it against a directly measured rank. We note that the
earlier, incorrect ceiling $3.5/t_{\max}$ would have implied $3\times(1+2\times7)
=45$ basis functions, which cannot explain a rank ceiling of $33$; the corrected
ceiling closes the chain and the incorrect one does not.

The closure is exact only at saturation. The intermediate ranks $5,15,25$ at
$L=1,2,3$ are not a single $3\times(1+2m)$ formula, because the reachable
harmonic set grows with depth and differs across the three observables at low
depth. At $L=1$ only the integer harmonics $\{0,1,2\}/t_{\max}$ are reachable and
they are split across observables ($\Zexp{0},\Zexp{1}$ each reach $\{1\}$,
$\Zexp{2}$ reaches $\{0,2\}$), so the per-observable block ranks are $2,2,3$ and
support counting bounds the stacked rank by $2+2+3=7$; two cross-observable
linear dependencies among the shared low-depth parameter directions reduce this
to the measured stacked rank of $5$. The qubit-3 half-integer harmonics become
reachable through the entangler only at $L\ge2$, the full band is reached on all
three observables at $L=4$ (with the cost observable $\Zexp{0}$ lagging the
others at $L=3$), and the every-initialisation rank saturation at $33$ completes
one layer later at $L=5$ ($32.7$ on average at $L=4$). The corrected ceiling
$f_{\max}$ is exact at every depth (there is never power above it); it is the
filling of the band, not the ceiling, that grows with depth.

\subsection{The trained loss is flat once the rank has saturated}
\label{ssec:capcurve}

The structural diagnostics predict an empirical signature: trained loss should
improve with depth up to the point where the Jacobian rank saturates
($L\approx4$--$5$), then stop improving. Table~\ref{tab:capcurve} is consistent
with this. At a matched $800$-iteration SPSA budget the mean final loss falls
from $L=1$ to $L=5$ and is flat from $L=5$ to $L=8$ ($1.08\times10^{3}$ against
$1.13\times10^{3}$). We report standard errors over the three seeds because they
qualify the strength of the claim. The dispersion is large at shallow depth
(standard errors of $2131,1581,566$ at $L=1,2,3$) and small once the loss has
plateaued ($86,52$ at $L=5,8$). The improvement from $L=3$ ($1957\pm566$) to
$L=5$ ($1080\pm86$) is therefore only about $1.5$ standard errors and the
per-seed ranges overlap, and $L=4$, where the rank actually saturates, was not
in the trained grid. The data are thus consistent with the loss plateauing where
the rank saturates, but the precise onset is not statistically resolved by three
seeds; we state the coincidence as consistency, not as an exactly located
transition. The robust part of the signal, monotone improvement over
$L=1$--$3$ and a flat plateau over $L=5$--$8$, matches the capacity picture.

\begin{table}[htbp]
  \centering
  \caption{Mean final loss (with standard error over $3$ seeds) versus depth at a
    matched $800$-iteration SPSA budget. Loss improves with depth up to $L=5$ and
    is then flat, consistent with the Jacobian-rank saturation of
    Table~\ref{tab:jacobian}; the shallow-depth spread is large, so the onset is
    reported as consistency rather than an exactly located transition.
    \textbf{Source:}
    \texttt{results/r1/train\_spsa\_b1\_e3\_cap\_L\{1,2,3,5,8\}.json} and
    \texttt{results/r1/r1\_corrections\_analysis.json}
    (\texttt{capacity\_dispersion}).}
  \label{tab:capcurve}
  \begin{tabular}{lcc}
    \toprule
    $L$ & Params & Mean final loss (SEM) \\
    \midrule
    $1$ & $15$  & $1.04\times10^{4}\ (2.1\times10^{3})$ \\
    $2$ & $30$  & $3.87\times10^{3}\ (1.6\times10^{3})$ \\
    $3$ & $45$  & $1.96\times10^{3}\ (5.7\times10^{2})$ \\
    $5$ & $75$  & $1.08\times10^{3}\ (8.6\times10^{1})$ \\
    $8$ & $120$ & $1.13\times10^{3}\ (5.2\times10^{1})$ \\
    \bottomrule
  \end{tabular}
\end{table}

\begin{figure}[htbp]
  \centering
  \includegraphics[width=0.9\linewidth]{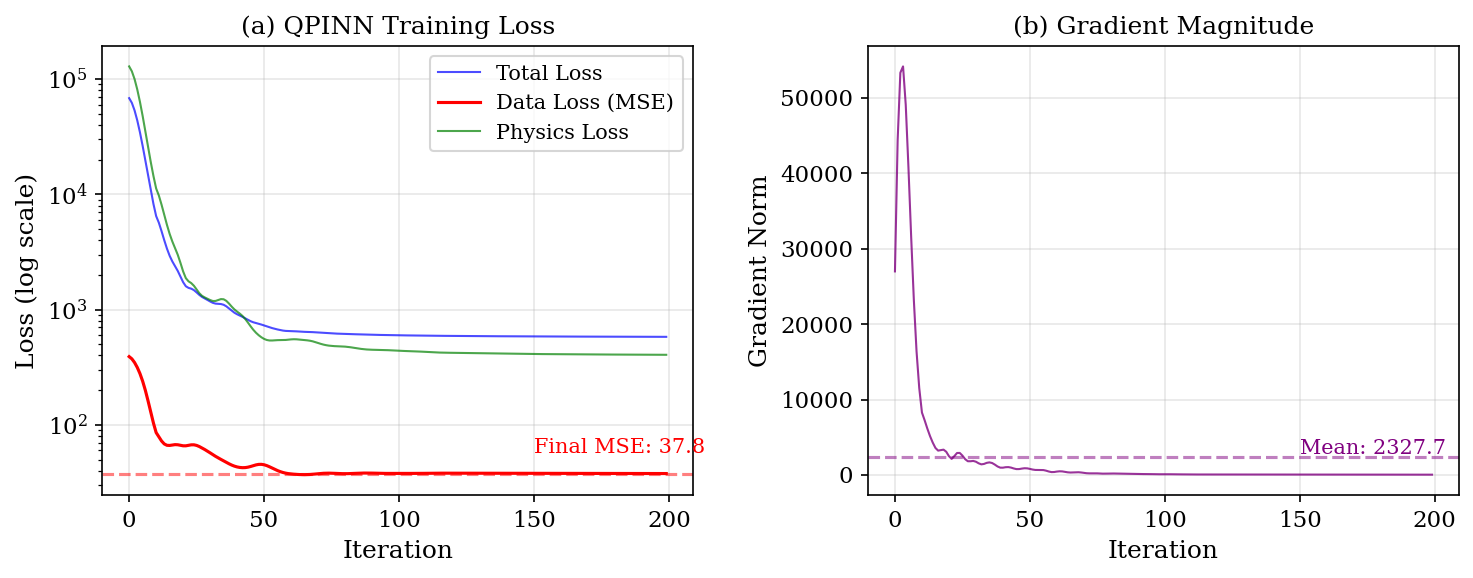}
  \caption{Representative QPINN training dynamics: the loss oscillates and
    plateaus while the gradient remains large throughout, the signature of an
    under-parameterised fit rather than of a barren plateau. (Reproduced from the
    setup of~\cite{pandey2604}.)}
  \label{fig:training}
\end{figure}

The training dynamics themselves (Figure~\ref{fig:training}) show the
oscillatory, high-across-seed-variance behaviour of an under-parameterised fit
with a live gradient, not the flat, stalled signature of a vanished gradient.

\section{Discussion}
\label{sec:discussion}

\subsection{Correcting our earlier barren-plateau diagnosis}
\label{ssec:walkback}

An earlier, unpublished version of this analysis reported that the QPINN was not
on a barren plateau on the basis that its measured gradient norms remained in the
range $10^{3}$--$10^{4}$ throughout training, four to five orders of magnitude
above the McClean threshold. That argument reached a defensible conclusion for
the wrong reason, and we correct the measurement here. The gradient norm of the
time-averaged physics loss is not a McClean-comparable quantity: it carries the
arbitrary scale of the unbounded, unnormalised loss, and averaging over the
encoding period before differentiating annihilates the very Fourier modes the
McClean bound concerns. Comparing an unnormalised
$\norm{\nabla\mathcal{L}}$ of order $10^{3}$ to a dimensionless $2^{-n}$
threshold is a category error.

The corrected measurement (Section~\ref{ssec:variance}) fixes this. It uses (a) a
bounded local cost $C_t\in[0,1]$, removing the arbitrary scale; (b) a fixed
circuit at a fixed time, so no time-averaging destroys the relevant modes; and
(c) exact parameter-shift gradients, removing any finite-difference artefact. On
this quantity the variance sits at the local-cost Haar/2-design floor
$2^{-2n}=3.9\times10^{-3}$, and on its live parameters it decays toward that
floor with depth rather than falling to zero. The earlier four-to-five-orders
cushion was an artefact of the unnormalised loss scale and does not survive
normalisation: after normalisation the variance is at the $2$-design floor, not
far above it. The conclusion (no vanished gradient) is unchanged; the magnitude
of the claim is much more modest, and the correct reading is that gradient
magnitude is not the binding constraint, capacity is.

We also do not repeat the earlier framing that presented this disambiguation as a
novel general-purpose diagnostic. The contribution here is narrower and more
defensible: for this specific circuit and task, four measurements agree that the
ceiling is capacity/expressivity, not a plateau.

\subsection{Why depth cannot help: encoding-limited expressivity}

The four diagnostics form a single mechanistic chain. The encoding fixes the
accessible frequency band at $2.5/t_{\max}$
(Section~\ref{ssec:fourier}); the band is narrower than the target, so the
reachable output functions span a fixed space of dimension $33$; the Jacobian
rank therefore saturates at $33$ once depth is sufficient to populate that space
(Sections~\ref{ssec:jacobian} and~\ref{ssec:closure}); the trained loss is then
flat at the same depth (Section~\ref{ssec:capcurve}); and throughout, the
gradient variance sits at the Haar floor and is large enough to train
(Section~\ref{ssec:variance}), with no optimiser doing better
(Section~\ref{ssec:optimiser}). The bottleneck is the fixed encoding, not the
number of trainable parameters and not the optimiser. This is a concrete
instance of the Schuld et al.~\cite{schuld2021effect} result that the encoding
sets the model's Fourier support: to fit a higher-bandwidth target one must
change the encoding (for example richer data re-uploading), not merely add
layers.

\subsection{Architectural, not quantum: the reservoir and ESN comparison}
\label{ssec:architectural}

On the same Lorenz task, a fixed quantum reservoir with a closed-form ridge
readout and a purely classical echo-state network (ESN) both outperform the
variational QPINN by a wide margin, and in a fair comparison the tuned classical
ESN is the strongest of the three~\cite{pandey2604,pandey2607b}. It
would be tempting to read the reservoir's advantage as a quantum effect; the
diagnosis here shows it is not. The relevant contrast is architectural: a fixed,
high-dimensional random feature map with a closed-form linear readout against a
small, band-limited, end-to-end-trained circuit. The former avoids the capacity
ceiling by supplying a rich fixed feature basis and solving the readout in closed
form (Figure~\ref{fig:trajectory}); the latter is limited by the fixed encoding's
narrow Fourier support and by
its handful of trainable parameters. Because both the classical ESN and the
quantum reservoir enjoy the architectural advantage, the advantage is not quantum
mechanical. Our companion conditioning
analysis~\cite{pandey2607} makes the same point from the
numerical-stability side: the fixed random feature map has a bounded condition
number where the trained alternative does not, so the architectural gap is
visible in the conditioning of the readout problem independently of any quantum
consideration.

\begin{figure}[htbp]
  \centering
  \includegraphics[width=0.9\linewidth]{fig1_trajectory.png}
  \caption{The fixed high-dimensional reservoir tracks the chaotic trajectory
    that the band-limited variational circuit cannot. Quantum-reservoir (QRC)
    predictions with a closed-form ridge readout, overlaid on the reference
    Lorenz-63 solution. (a) $3$D attractor view: the QRC predictions (red)
    follow the reference orbit. (b) Component-wise time series (offset for
    clarity): QRC $x(t)$ and $y(t)$ track the reference to within marker width.
    The fixed random feature basis supplies enough capacity to reconstruct the
    orbit, whereas the fixed-encoding variational circuit is capped by its narrow
    Fourier support (Section~\ref{ssec:fourier}); the contrast is architectural
    rather than quantum.}
  \label{fig:trajectory}
\end{figure}

The summary is therefore modest and specific. We do not claim a quantum
advantage; we give a mechanistic explanation for a widely observed
variational-circuit failure, and we locate the observed performance differences
in architecture rather than in quantum physics.

\subsection{Scope and limitations}

We deliberately do not make near-term-hardware claims. The study is a four-qubit,
exactly simulated diagnosis; the value of the small scale is that every
diagnostic is exact, so the barren-plateau question can be settled without
sampling artefacts, not that the result is a device demonstration. Three
limitations follow. First, the barren-plateau statement is a scaling claim in
the qubit number, which a single $n=4$ experiment cannot settle in the
asymptotic sense; what we establish is that at this width the gradient is at the
Haar floor and is large enough to train, so the observed failure is not
gradient-magnitude-limited. Second, the encoding-set frequency ceiling is a
property of this encoding; a different encoding (denser data re-uploading,
learnable encoding frequencies) would move the ceiling, and testing that is the
natural next experiment. Third, we have shown that depth does not help this
architecture on this target; we have not shown that no small variational circuit
can fit chaotic dynamics, only that fitting requires widening the accessible
band, which is an encoding choice rather than a depth choice.

\section{Conclusion}

The failure of a four-qubit variational quantum circuit to fit the Lorenz-63
attractor is not a barren plateau. A McClean-comparable, fixed-circuit,
exact-parameter-shift gradient variance sits at the local-cost Haar floor
$2^{-2n}=3.9\times10^{-3}$ and, on its live parameters, decays toward that floor
with depth rather than collapsing to zero; three optimiser families reach the
same order of magnitude of loss at matched budget; the output-Jacobian rank
saturates at $33$ from five layers on; and the accessible output frequency band
is fixed by the encoding at $2.5/t_{\max}\approx0.83$~Hz, a factor of about
$4.4$ below even the narrowest target component, at every depth. The corrected
band has dimension $3\times(1+2\times5)=33$, exactly the measured rank ceiling,
tying the Fourier and Jacobian diagnostics into one number. These measurements
agree on a single mechanism: an architectural capacity ceiling set by the fixed
encoding, not a trainability plateau. We correct our earlier, unnormalised
gradient-norm diagnosis, and we place the observed advantage of fixed reservoirs
and classical echo-state networks on this task in architecture rather than in
quantum mechanics.

\ack
The author used Claude (Anthropic) to assist with running and optimizing the
simulation code. All technical content, results, and conclusions are the
author's own. This work received no external funding. The author declares no
conflicts of interest.

\section*{Data and code availability}
\label{sec:availability}

All diagnostic code and result files (with recorded git revision and per-file
SHA-256 provenance) are available at
\url{https://github.com/pandey-tushar/Quantum_Chaos_solver}. The diagnostics of
Section~\ref{sec:results} are produced by \texttt{scripts/r1\_experiments.py};
the corrected live/dead variance split and capacity dispersion are computed by
\texttt{scripts/analyze\_r1\_corrections.py}, and the corrected Fourier ceiling
by \texttt{scripts/verify\_fourier\_ceiling.py}. The result files are named in
each table caption.



\begin{thebibliography}{99}

\bibitem{mcclean2018}
J.~R. McClean, S.~Boixo, V.~N. Smelyanskiy, R.~Babbush, and H.~Neven,
Barren plateaus in quantum neural network training landscapes,
\textit{Nat.\ Commun.} \textbf{9}, 4812 (2018).

\bibitem{cerezo2021}
M.~Cerezo \textit{et al.},
Variational quantum algorithms,
\textit{Nat.\ Rev.\ Phys.} \textbf{3}, 625--644 (2021).

\bibitem{cerezo2021costlocality}
M.~Cerezo, A.~Sone, T.~Volkoff, L.~Cincio, and P.~J. Coles,
Cost function dependent barren plateaus in shallow parametrized quantum
circuits,
\textit{Nat.\ Commun.} \textbf{12}, 1791 (2021).

\bibitem{uvarov2021locality}
A.~V. Uvarov and J.~D. Biamonte,
On barren plateaus and cost function locality in variational quantum algorithms,
\textit{J.\ Phys.\ A: Math.\ Theor.} \textbf{54}, 245301 (2021).

\bibitem{ortiz2021entanglement}
C.~Ortiz Marrero, M.~Kieferov\'a, and N.~Wiebe,
Entanglement-induced barren plateaus,
\textit{PRX Quantum} \textbf{2}, 040316 (2021).

\bibitem{qi2026tensorhyper}
J.~Qi, C.-H.~H. Yang, P.-Y. Chen, and M.-H. Hsieh,
TensorHyper-VQC: a tensor-train-guided hypernetwork for robust and scalable
variational quantum computing,
\textit{npj Quantum Inf.} \textbf{12}, 70 (2026).

\bibitem{pandey2604}
T.~Pandey,
Quantum reservoir computing versus variational quantum physics-informed
networks for chaotic forecasting,
arXiv:2604.23743.

\bibitem{lorenz1963}
E.~N. Lorenz,
Deterministic nonperiodic flow,
\textit{J.\ Atmos.\ Sci.} \textbf{20}, 130--141 (1963).

\bibitem{raissi2019}
M.~Raissi, P.~Perdikaris, and G.~E. Karniadakis,
Physics-informed neural networks,
\textit{J.\ Comput.\ Phys.} \textbf{378}, 686--707 (2019).

\bibitem{qcpinn2025}
A.~Pattnaik \textit{et al.},
Quantum-classical physics-informed neural networks for solving nonlinear
differential equations,
arXiv:2503.16678 (2025).

\bibitem{leung2022qpinn}
Y.-C. Leung \textit{et al.},
Quantum physics-informed neural networks for solving differential equations,
arXiv:2212.00360 (2022).

\bibitem{schuld2021effect}
M.~Schuld, R.~Sweke, and J.~J. Meyer,
Effect of data encoding on the expressive power of variational
quantum-machine-learning models,
\textit{Phys.\ Rev.\ A} \textbf{103}, 032430 (2021).

\bibitem{jaeger2004harnessing}
H.~Jaeger and H.~Haas,
Harnessing nonlinearity: predicting chaotic systems and saving energy in
wireless communication,
\textit{Science} \textbf{304}, 78--80 (2004).

\bibitem{pathak2018model}
J.~Pathak, B.~Hunt, M.~Girvan, Z.~Lu, and E.~Ott,
Model-free prediction of large spatiotemporally chaotic systems from data,
\textit{Phys.\ Rev.\ Lett.} \textbf{120}, 024102 (2018).

\bibitem{fujii2017}
K.~Fujii and K.~Nakajima,
Harnessing disordered-ensemble quantum dynamics for machine learning,
\textit{Phys.\ Rev.\ Applied} \textbf{8}, 024030 (2017).

\bibitem{mujal2021}
P.~Mujal \textit{et al.},
Opportunities in quantum reservoir computing and extreme learning machines,
\textit{Adv.\ Quantum Technol.} \textbf{4}, 2100027 (2021).

\bibitem{nakajima2019boosting}
K.~Nakajima, K.~Fujii, M.~Negoro, K.~Mitarai, and M.~Kitagawa,
Boosting computational power through spatial multiplexing in quantum reservoir
computing,
\textit{Phys.\ Rev.\ Applied} \textbf{11}, 034021 (2019).

\bibitem{ahmed2025robust}
S.~Ahmed, L.~Tennie, and L.~Magri,
Robust quantum reservoir computing for chaotic dynamics,
\textit{Proc.\ R.\ Soc.\ A} \textbf{481}, 20250550 (2025).

\bibitem{pandey2607}
T.~Pandey,
A quantum reservoir architecture for chaotic forecasting and a test of
whether its high dimension helps,
arXiv:2607.07978 (2026).

\bibitem{pandey2607b}
T.~Pandey,
When classical baselines are tuned as carefully as the quantum model, does
quantum reservoir computing still win?,
arXiv:2607.09905 (2026).

\end{thebibliography}
\end{document}